\documentclass[prl,twocolumn,groupedaddress,showpacs,byrevtex]{revtex4}
\usepackage[]{umlaut}
\usepackage[]{amssymb}
\usepackage[]{graphicx}

\begin{document}

\bibliographystyle{apsrev}

\title{Temperature dependence of the Kondo resonance and its satellites
in CeCu$_2$Si$_2$}  
\author{F. Reinert}
\email[corresponding author. Email: ]{friedel@mx.uni-saarland.de}
\author{D. Ehm}
\author{S. Schmidt}
\author{G. Nicolay}
\author{S. H\"ufner}
\affiliation{Universit\"at des Saarlandes, Fachrichtung 7.2 ---
  Experimentalphysik, 66041 Saarbr\"ucken, Germany} 
\author{J. Kroha}
\affiliation{Institut f\"ur Theorie der Kondensierten Materie,
Universit\"at Karlsruhe, Engesserstr.\ 7, 76128 Karlsruhe, Germany}
\author{O. Trovarelli}
\author{C. Geibel}
\affiliation{Max-Planck Institute for Chemical Physics of Solids,
  N\"othnitzer Str. 40, 01187 Dresden, Germany}

\date{\today}

\begin{abstract}
We present high-resolution photoemission 
spectroscopy studies on the
Kondo resonance of the strongly-correlated Ce system CeCu$_2$Si$_2$.
Exploiting the thermal broadening of the Fermi edge we analyze position,
spectral weight, and temperature dependence of the low-energy
4f spectral features, whose major weight lies above the 
Fermi level $E_F$. We also present theoretical predictions 
based on the single-impurity Anderson model using an extended
non-crossing approximation (NCA), including all
spin-orbit and crystal field splittings of the 4f states. 
The excellent agreement between theory and experiment provides strong
evidence that the spectral properties of CeCu$_2$Si$_2$ can be 
described by single-impurity Kondo
physics down to $T \approx 5$~K.
\end{abstract}
\maketitle

Many of the salient properties of intermetallic rare earth compounds
originate from strong electronic correlations in the
rare earth 4f orbitals. The hybridization with the conduction electron 
continuum leads to complex low-temperature behavior, 
summarized by the term Kondo scenario \cite{hewson93}. 
It includes quenching of the 4f magnetic moments below the Kondo 
temperature $T_K$, a strong enhancement of thermodynamic 
quantities like the magnetic susceptibility and the specific 
heat. The latter is ascribed to an enhanced electron
density of states (DOS) at the Fermi energy $E_F$, the Kondo resonance (KR).
In lattice systems, below a so-called lattice coherence temperature $T^*$,
the transport properties \cite{fulde95}  
can often be described by the assumption 
of {\it propagating} quasiparticles 
with an effective mass enhancement
of $m^*_e/m^{\phantom{*}}_e = 10^2$ to $10^3$. The coherence temperature $T^*$ can be in the 
same range as $T_K$ \cite{knebel96}. Several of these heavy Fermion systems
\cite{gschneidner} undergo a 
magnetic ordering transition at low temperature $T$, become
paramagnetic insulators (Kondo insulators) or even
superconducting, like CeCu$_2$Si$_2$ \cite{steglich79},
where $T_K \approx 4.5$ K to 10 K \cite{bredl84,horn81} 
and $T^* \approx 10$ K \cite{knebel96}. 
In order to understand the interplay between local moment quenching
and lattice coherence and ordering effects it is essential to know
to what extent the system is described by local Kondo physics. 
However, it has been notoriously difficult to directly investigate 
the KR experimentally. In fact, the Kondo scenario as the origin of
the 4f DOS enhancement in Ce-based compounds has repeatedly been 
disputed in the literature (see \cite{arko_gschneidner99} and
references therein), mainly because the enhancement can still be seen
at temperatures almost two orders of magnitude above $T_K$.

There exist three established, spectroscopic methods to investigate 
the spectral function close to the Fermi energy,
photoemission (PES), inverse photoemission (IPES) and scanning
tunneling spectroscopy (STS).  
In the case of Ce compounds the 4f occupation is close to unity
(4f$^1$), the Kondo temperature lies in the range
$T_K \approx 1, \dots , 1000$ K \cite{gschneidner}, and the KR 
has its maximum {\em above\/} $E_F$. 
The difficulty in observing the KR lies in the limited IPES
energy resolution of $O(100 \,{\rm meV})$, 
whereas PES has, in general, merely access to the tail of the KR below $E_F$ 
and STS \cite{li98} probes the local 4f spectrum only
indirectly by tunneling into the conduction states \cite{ujsaghy00}.

In this Letter we present a careful analysis of high-resolution
PES data close to the Fermi level, which allows to acquire
information of the spectral function above the Fermi level. 
We restrict ourselves here to the heavy fermion compound
CeCu$_2$Si$_2$, 
although experimental method and data analysis \cite{kumigashira99}
have also been successfully applied to other Ce intermetallics \cite{ehm_diss}.
A comparison with calculations of the 4f spectral
function, based on the single-impurity Anderson model (SIAM) including 
spin-orbit (SO) as well as crystal field (CF) splitting of the local orbitals,
shows striking agreement between experiment and theory.
In particular, we explain the
temperature dependence of the spectral features up to $T=200$~K.

The PES experiments have been performed using a SCIENTA SES~200
analyzer
and a monochromatized GAMMADATA VUV-lamp at photon energies of
$h\nu=21.2$~eV (He~I) and $h\nu=40.8$~eV (He~II). 
A description of the experimental setup and the calibration of the
experimental parameters for the high-resolution 
measurements, like $E_F$, sample temperature $T$, and energy resolution $\Delta E$, 
can be found in more detail in Ref.\ \cite{cuagau_reinert01}. 
The poly-crystalline samples were cleaved {\em in situ} at low
$T$ to prepare clean surfaces. Because of the
high surface reactivity of rare-earth compounds
\cite{yic_reinert98} the duration of a 
measurement was kept less than about 12~h.

\begin{figure}[tb]
  \begin{center}
    \includegraphics[angle = -90, width = 8.3cm]{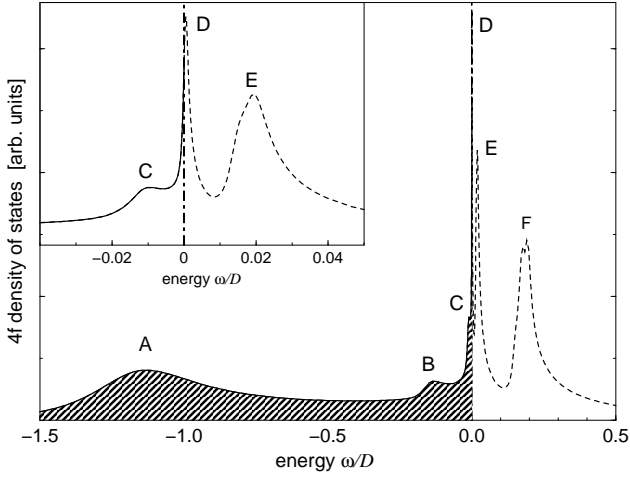}
    \caption[nca_dos]{Theoretical 4f spectral function from
      calculations based on the SIAM using the NCA for $T=11$ K and
      model parameters as in Fig.\ \ref{fig:tempdep}. The hatched
      region is the occupied part of the spectrum. Energies are given 
      in units of the half band width $D=E_F$. The inset shows
      the near--$E_F$ region. The spectral features {\sf A} -- {\sf F} 
      are explained in the text. The two-electron state $f^2$ lies 
      far outside the displayed energy range at 
      $\approx U+\epsilon_{f1}$.}
    \label{fig:NCA_DOS}
  \end{center}
\end{figure}

\begin{figure}[tb]
  \begin{center}
    \includegraphics[width = 8.2cm]{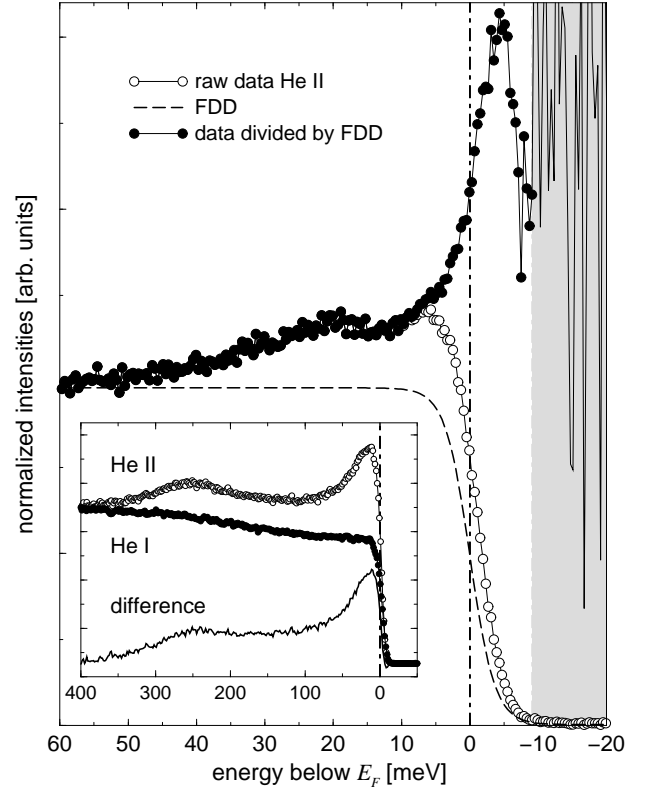}    
    \caption{Photoemission spectrum of CeCu$_2$Si$_2$ ($h\nu=40.8$~eV,
      He~II). 
      Open circles represent the raw data close to the Fermi level,
      including the tail of the KR and the low-energy
      CF excitations at approximately 21~meV  
      ($\Delta E = 6$~meV, $T=11$~K). Filled circles represent the 
      raw data divided by the experimentally broadened FDD (see text).
      The shaded area marks the unreliable
      spectral range above $5k_BT$. The inset shows
      an extended energy range (measured with $\Delta E\sim 13$~meV, 
      $T=17$~K) including the spin-orbit (J=7/2) excitation at
      $\approx 270$~meV below $E_F$ and the
      non-4f background as determined using He~I ($h\nu=21.2$~eV) radiation.}
    \label{fig:ex_PES}
  \end{center}
\end{figure}

In CeCu$_2$Si$_2$ the 7 spin--degenerate 4f levels are split by SO coupling 
into a total angular momentum $J=5/2$ sextet and an excited
$J=7/2$ octet. These are, in turn, CF split into 3 and 4
Kramers degenerate doublets, respectively. Thus, the
single-impurity system is described by the multi--orbital 
Anderson Hamiltonian 
\begin{eqnarray}
H &=& H_0
   +  \sum _{m\sigma} \varepsilon _{f m} 
      f^{\dagger}_{m\sigma}f^{\phantom{\dagger}}_{m\sigma}  \nonumber  
   +  \sum _{\vec km\sigma} \bigl( 
      V_{\vec km} c^{\dagger}_{\vec k\sigma}f^{\phantom{\dagger}}_{m\sigma} + 
      {\rm h.c.} \bigr) \\
  &+&  \frac{U}{2} \sum _{(m\sigma )\neq(m ' \sigma ')}
  f^{\dagger}_{m\sigma}f^{\phantom{\dagger}}_{m\sigma}
  f^{\dagger}_{m '\sigma '}f^{\phantom{\dagger}}_{m '\sigma '} \ ,
\label{Amodel}
\end{eqnarray}
where $H_0=
      \sum _{\vec k\sigma} 
      \varepsilon _{\vec k} c^{\dagger}_{\vec k\sigma}
                     c^{\phantom{\dagger}}_{\vec k\sigma}$
describes the conduction band with dispersion $\varepsilon _{\vec k}$ and creation 
operators $c^{\dagger}_{\vec k\sigma}$ for electrons with spin $\sigma$.
$\varepsilon _{fm} < E_F$, $m=1,\dots ,7$, are the SO and CF split 4f
single-particle levels with the corresponding creation operators
$f^{\dagger}_{m\sigma}$.
The hybridization matrix elements $V_{\vec km}$ lead to an effective 
coupling matrix between 4f states, $\Gamma _{mn}=\pi \sum_{\vec k}
V^*_{m\vec k} A_{\vec k}(\omega ) V_{\vec k n}$, where 
$A_{\vec k}(\omega )$ is the conduction electron spectral function.
The local Coulomb repulsion, known from IPES
\cite{grioni97,kanai99}, is substantially larger 
than $|\varepsilon_{f1}|$ 
and, hence, effectively suppresses any double occupancy 
of the Ce 4f levels ($U\to \infty$). 
In the multi--orbital system the Kondo temperature is given by 
$T_K \approx \sqrt{2 J E_F}\;  {\rm exp}[-1/(2N(0)J)]$, where $N(0)$ is the
conduction electron DOS at the Fermi level and the Kondo coupling
$J$ is obtained from a Schrieffer--Wolff transformation including
the CF and SO states as
\begin{eqnarray}
J= \frac{|\sum_{\vec k} V_{\vec k 1} |^2}{\Bigl| \varepsilon _{f1} +
\sum_{m > 1} \frac{|\sum _{\vec k} V_{\vec k m}|^2}{\varepsilon_{fm}-
\varepsilon_{f1} }\Bigr| }.
\end{eqnarray}
It is seen that $J$, and hence $T_K$, is enhanced due to coupling to 
the CF and SO split 4f states.

To achieve a detailed comparison with experimental results,
we have calculated the 4f spectral function of the model 
Eq.~(\ref{Amodel}) using the non-crossing approximation (NCA) 
\cite{bickers87,costi96} including all CF and SO excitations. 
The NCA is known to reliably describe the low-temperature scale $T_K$ and the
position, the spectral weight and the life-time broadening  
of the peaks in the local spectral density down to $T \approx 0.1\, T_K$
\cite{bickers87}. The spectrum has generically six 
distinct features as shown in Fig.~\ref{fig:NCA_DOS} ({\sf A}--{\sf F}). 
They can be understood as follows: At low $T$ the occupation of the 
lowest 4f level is close to unity ($n_{f1}\lesssim 1$), while
all other single-particle 4f states are essentially empty 
($n_{fm} \approx 0$, $m=2,\dots,7$) because of the strong Coulomb 
repulsion $U$. 
Hence, the broad 4f$^1$ $\to$ 4f$^0$ ionization peak ({\sf A}) 
with a full width at half maximum (FWHM) of $\Gamma \approx \sum_m \Gamma _{1m}$
corresponds to the lowest single-particle
level $\varepsilon _{f1}$. Resonant spin flip scattering of electrons at the 
Fermi energy induces the narrow Kondo resonance ({\sf D}) of width $\sim k_BT_K$,
somewhat shifted above $E_F$ due to level repulsion 
from the single-particle levels $\varepsilon _{fm} <E_F$.
The SO and the CF satellite peaks appear in pairs ({\sf B}, {\sf F}) and
({\sf C}, {\sf E}), respectively. They arise from virtual transitions from 
the ground state into the (empty) excited 
SO ({\sf F}) and CF ({\sf E}) states and vice versa
({\sf B} and {\sf C}). The positions
of the satellite peak pairs are, therefore, approximately symmetrical about $E_F$.
However, while the features above $E_F$ have significant spectral weight,
those below $E_F$ appear merely as weak shoulders. This is because the
transition probabilities carry a detailed balance factor $w = n^i(1-n^f)$,
where $n^i$ ($n^f$) is the occupation number of the 4f orbital in the
initial (final) state, i.e. $w$ is large for the excitations {\sf E}, {\sf F},
but small for the transitions {\sf B}, {\sf C}. 

As $n^i$, $n^f$ are controlled both by $T$ and $U$,
the CF and SO satellites are signatures of strong correlations and are
$T$-dependent. All these features are
well described by the NCA (Fig.\ \ref{fig:NCA_DOS}). 
We note in passing that
similar spectra should be observed in multilevel quantum dots
in the Kondo regime.

\begin{figure}[tb]
  \begin{center}
    \includegraphics[width = 7.7cm]{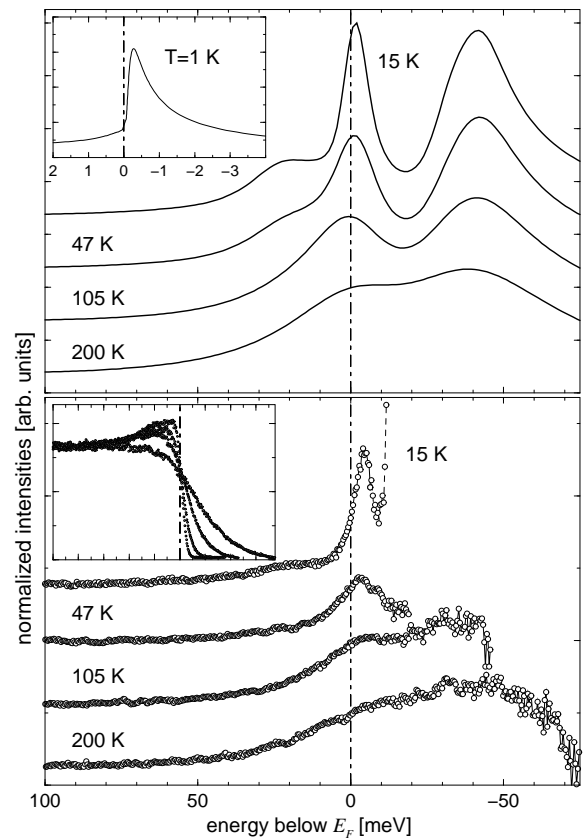}
    \caption[TEMPERATURE DEPENDENCE]{Upper panel: 
      Theoretical $T$ dependence of the 4f
      spectral function of CeCu$_2$Si$_2$. 
      The inset shows the calculated spectrum below $T_K$ (see text).
      Used model parameters: $\epsilon_{f1}=-2.4$~eV, $E_F=4$~eV, CF
      splittings of the $J=5/2$ sextet $\Delta_{CF}=30$~meV and $36$~meV, SO
      splitting $\Delta_{SO}=270$~meV, hybridization $V=200$~meV.
      Lower panel: Photoemission spectra for $T=15$~K, $47$~K, $105$~K, and $200$~K. 
      The experimental spectra, divided by the FDD, are clipped on
      the high energy side at $\approx5k_BT$.  
      The inset shows the data on the same energy scale
      prior to the division. All spectra are normalized to the same
      intensity at $\approx100$~meV and offset for clarity. }
    \label{fig:tempdep}
  \end{center}
\end{figure}

In order to identify the low-energy features in experimental
spectra as due to the physics of the single-impurity Anderson model
(Eq.\ \ref{Amodel}), it is highly desirable to gain access to the 
sharp resonances {\sf D}, {\sf E} above $E_F$.
A typical high-resolution spectrum on CeCu$_2$Si$_2$ is shown in
Fig.~\ref{fig:ex_PES}, measured at $T=11$~K with an energy resolution
of $\Delta E=6$~meV. 
The ionization peak {\sf A} (not shown) appears at a binding
energy of $\approx2$~eV below $E_F$, the SO satellite {\sf B} at
$\approx270$~meV. Fig.~\ref{fig:ex_PES} shows the near-$E_F$ region 
with the tail of the KR and a distinct spectral feature at about $21$~meV, 
which can be assigned to the CF splitting of the $J=5/2$ 4f levels
\cite{goremychkin93}. In this plot the spectral intensity  
at energies of a few meV above $E_F$ and higher is completely suppressed 
by the Fermi-Dirac distribution (FDD). 
IPES experiments performed on similar systems do not resolve the 
KR and its satellite excitations expected above $E_F$, but show only one broad
structure near $E_F$ and a second one at much
higher energies, corresponding to the doubly occupied IPES final 
state 4f$^1 \rightarrow$ 4f$^2$ \cite{grioni97,kanai99}.

Greber {\em et al.} \cite{greber97} have demonstrated that a careful 
analysis of PES data based on a division by the respective FDD allows to
investigate the spectral function up to energies of approximately
$5k_BT$ {\em above} $E_F$, provided that the noise of the data and
the experimental broadening are small enough.  
We applied this procedure to the spectrum in Fig.~\ref{fig:ex_PES}, 
where we divided the raw spectrum by the FDD at $T=11$~K convoluted with 
the spectrometer function, described by 
a Gaussian of full width at half maximum (FWHM) $\Delta E=6$~meV. Obviously there appears a
narrow peak with a FWHM of 6 meV and a maximum at about $3$~meV above $E_F$. 
This is sufficiently below $5k_BT\approx7$~meV at which energy the 
scatter of the data gets large due to the exponential increase of the 
inverse FDD. The spectrum of the CF satellite at $\approx21$~meV below
$E_F$ remains unchanged by this procedure. It should be mentioned that
the described procedure is not exact and therefore can lead to
artifacts (of the order of $\Delta E$) if the temperature broadening is
comparable with the energy resolution. However, in the present case
this is only relevant for the
11~K-spectrum which appears slightly shifted away from $E_F$.  

In the following we analyze the $T$ dependence of the near--$E_F$ 
spectral features. Fig.\ \ref{fig:tempdep}, lower panel, shows the
experimental data at several temperatures, divided by the
broadened FDD as described above. 
The upper panel of Fig.\ \ref{fig:tempdep}
displays the NCA spectral functions for the same temperatures.
The model parameters, i.e. the single-particle energies $\epsilon_{fm}$ 
and the hybridization strengths $\Gamma_{m m'}$, have been taken from
independent experimental measurements where possible 
\cite{goremychkin93,kang90}
and have only slightly been adjusted (see caption of Fig.\ \ref{fig:tempdep})
to get optimal quantitative agreement with our experimental results. 
The experimentally observed temperature
dependence of the spectra is accurately described by the theoretical 
simulations, and a comparison allows to identify the origin of the
various spectral features: 
As already seen in Fig.\ \ref{fig:ex_PES} at low $T$ ($T=15$~K) 
the KR at $E_F$ 
appears as a narrow line with a FWHM of about 6~meV. Due to the given
energy resolution of $\Delta E=6$~meV the intrinsic
linewidth of the Kondo resonance must be much smaller than $\Delta E$
at low temperatures. In order to estimate $T_K$ from the
data we have determined the model parameters to fit the experimental results
at their respective $T$ and then calculated the spectrum at $T=0.1\; T_K \approx
1$ K (inset of Fig.\ \ref{fig:ex_PES}, upper panel), 
i.e. close to the unitarity limit which is not directly reachable 
in the experiment. $T_K$ was then determined from the peak width of this
theoretical low-$T$ spectrum as $T_K \approx 6$ K.   
This result is in remarkable agreement with thermodynamic bulk measurements
\cite{steglich79,bredl84,horn81},  
considering especially that the spectral function --- and therewith
$T_K$ --- as measured by surface sensitive PES might be
modified by surface effects, as pointed out by several authors
\cite{yic_reinert98,garnier97,sekiyama00}. Towards higher $T$,
the line width of the KR increases, while the maximum intensity becomes
smaller and the distinction of KR and CF satellite is successively 
smeared out. At $T=200$ K and above the thermal broadening of the FDD is
large enough for PES to have access to the CF excitations at $\sim 40$ meV.
At these elevated $T$ the KR has disappeared as a separate peak, 
and the enhanced 4f DOS near $E_F$ is rather due to the CF excitations
which are broadened and positioned somewhat above $E_F$,
as seen in Fig.\ \ref{fig:tempdep}. Thus, the persistance of an enhanced
DOS at the Fermi level even at room temperature, despite a substantially 
lower $T_K$, is naturally explained within the SIAM in combination with
CF excited states. 

In conclusion we have demonstrated that high resolution photoemission
spectroscopy gives detailed access to the spectral function of heavy
Fermion Ce compounds
and allows to investigate the temperature dependence of the Kondo resonance. 
Within an energy range of $5k_BT$ above the Fermi level several structures 
can be resolved, whose origin and temperature dependence are consistently 
described within the single-impurity Anderson model.
This excellent agreement provides strong evidence that the 4f
photoemission spectra of CeCu$_2$Si$_2$ are an immediate consequence
of single-impurity Kondo physics, although the Kondo local moment 
quenching and lattice coherence effects occur on similar 
energy scales ($T^*\approx T_K$). In particular, the Kondo resonance could be
unambiguously identified, with a Kondo temperature in reasonable agreement
with estimates from thermodynamic measurements. The persistence of
an enhanced density of states even at room temperature is explained
by the presence of crystal field excited states close to the Fermi
energy.

We wish to thank P. W\"olfle for stimulating discussions.
This work was supported by the Deutsche Forschungsgemeinschaft (grant
nos. HU149-19-1, HU149-17-4) and by Sonderforschungsbereiche
SFB 277 and SFB 195.

\end{document}